\documentstyle[12pt]{article}

\normalbaselines
\font\tbf = cmbx12

\pagebreak

\begin{document}
\indent
\vskip 1.5cm
\centerline{\tbf $Z_3$-GRADED  EXTERIOR  DIFFERENTIAL  CALCULUS}
\vskip 0.3cm
\centerline{\tbf AND  GAUGE  THEORIES  OF  HIGHER  ORDER} 
\vskip 1cm
\centerline{\tbf by}
\vskip 0.5cm
\centerline{\tbf Richard Kerner}
\vskip 1.5cm
\centerline{Laboratoire de Gravitation et Cosmologie Relativistes}
\vskip 0.2cm
\centerline{Universit\'e Pierre-et-Marie-Curie, CNRS - D0 769}
\vskip 0.3cm
\centerline{Tour 22, 4-\`eme \'etage, Bo\^{i}te 142}
\vskip 0.2cm
\centerline{4, Place Jussieu, 75005  Paris}
\vskip 0.2cm
\centerline{FRANCE}
\vskip 1cm
\indent
{\tbf Summary}. We present a possible generalization of the exterior differential 
calculus, based on the operator $d$ such that $d^3 = 0$, but $d^2 \neq 0$. 
The entities $d x^i$ and $d^2 x^k$ generate an associative algebra; we shall 
suppose that the products $dx^i dx^k $ are independent of $dx ^k dx^i$, while 
the {\it ternary} products will satisfy the relation: 
$dx^i dx^k dx^m = j dx^k dx^m dx^i = j^2 dx^m dx^i dx^k $ , complemented by
the relation $d x^i d^2 x^k = j d^2 x^k d x^i $, with $j := e^{\frac{2 \pi i}{3}}$.
\newline
\indent
We shall attribute grade 1 to the differentials $d x^i$ and grade 2 to the
"second differentials" $d^2 x^k$ ; under the associative multiplication law
the grades add up modulo 3.
\newline
\indent
We show how the notion of {\it covariant} derivation can be generalized with
a 1-form $A$ so that $D \Phi := d \Phi + A \Phi$ , and we give the expression
in local coordinates of the {\it curvature 3-form} defined as
$\Omega := d^2 A + d (A^2) + A dA + A^3$.
\newline
\indent
Finally, the introduction of notions of a scalar product and integration 
of the $Z_3$-graded exterior forms enables us to define variational principle
and to derive the differential equations satisfied by the 3-form $\Omega$.
The lagrangian obtained in this way contains the invariants of the ordinary 
gauge field tensor $F_{ik}$ and its covariant derivatives $D_i F_{km}$.
\newpage
\vskip 0.4cm
\indent
{\tbf 1. INTRODUCTION}
\vskip 0.4cm
\indent
The models of fundamental interactions and field theories based on the non-
commutative geometry have been the object of intense studies in past few
years (cf. refs. [1] to [6]). Most of the effort has been concentrated on
reproducing the Weinberg-Salam unified theory of electroweak interactions
by means of generalized gauge theories developed in non-commutative geometries
instead of fibre bundles, and with finite algebra of matrices replacing the
infinite algebra of functions on the manifold.
\newline
\indent
In most developed variants of this approach, the $Z_2$-{\it graded} matrix 
algebras have been used, whose $Z_2$-graded internal differential (defined as a 
graded commutator with a matrix whose square was equal to {\tbf 1}) was 
combined with the usual exterior differential forming a $Z_2$-graded Grassmann 
algebra. In the tensor products of these two algebras, i.e. in the algebra
of matrix-valued exterior forms, the $Z_2$-grades of the matrices were added 
(modulo $2$) to the $Z_2$-grades of the exterior forms, defining the $Z_2$-
grade of the whole object (cf. refs. [5] and [6]).
\newline
\indent
Here we would like to generalize this scheme to the case of $Z_3$-grading.
The resulting matrix algebra can be represented in the simplest case as
the algebra of $3 \times 3$ matrices with entries from an associative
algebra over complex numbers, on which a $Z_3$-graded commutator replaces 
the usual commutation and anti-commutational rules for the $Z_2$-graded
$2 \times 2$-matrices.
\newline
\indent
This leads naturally to the differential $d$ whose square $d^2$ is different
from $0$, but whose {\it cube} does vanish identically, $d^3 = 0$.
We show how such matrix algebra appears naturally as the algebra of linear
transformations of a $Z_3$-graded generalization of the Grassmann algebra.
We also introduce a $Z_3$-graded generalization of exterior algebra of
differential forms and show how the notions of connection and curvature can
be generalized, too, and how the lagrangians containing the invariants of
the curvature and its derivatives can be constructed.
\newline
\indent
Finally, we briefly discuss the resulting variational principle for the
gauge fields, leading to invariant equations of fourth order. Similar theories
with even higher order derivatives of the curvature and lagrangians depending
on their invariants can be obtained with a similar scheme based on $Z_N$-graded
differentials satisfying $d^N = 0$.

\vskip 0.5cm
{\tbf 2. $Z_3$-GRADED  ANALOG  OF GRASSMANN  ALGEBRA.}
\vskip 0.4cm
\indent
The cyclic group $Z_3$ can be represented in the complex plane by means of
the cubic roots of {\tbf 1} : let $j := e^{\frac{2 \pi i}{3}}$; then one has
$j^3 = 1$ and $j + j^2 + 1 = 0$; obviously, $j^n = j^{(n + 3)}$.
\newline
\indent
By analogy with the $Z_2$-graded Grassmann algebras spanned by the set of
anti-commuting generators, we may introduce an associative algebra spanned
by $N$ generators $\theta^A$, $ A,B = 1,2...N$, whose {\it binary} products
$\theta^A \theta^B$ will be considered as $N^2$  independent quantities, 
whereas we shall impose a {\it ternary} analog of the anti-commutation 
relations:
\begin{equation}
\theta^A \theta^B \theta^C = j  \theta^B \theta^C \theta^A = j^2 \theta^C
\theta^A \theta^B
\end{equation}
\indent
A more precise formulation is to say that the algebra in question is the
universal algebra defined by the above relations.
\newline
\indent
{\tbf Corollary} : The cube of any generator must vanish (because in this 
case the relation (1) amounts to $(\theta^A)^3 = j (\theta^A)^3 = 0$ ; all the 
monomials of order $4$ or higher are identically null (the proof that follows
makes use of the associativity of the postulated product and of the relation
$1$): (the low braces are there just to indicate to which triple of $\theta$'s
the circular permutation is being applied)
\begin{equation}
{\underbrace{\theta^A \theta^B \theta}}^C \theta^D = 
j \theta^B {\underbrace{\theta^C \theta^A \theta}}^D =
j^2 {\underbrace{\theta^B \theta^A \theta}}^D \theta^C = 
\theta^A {\underbrace{\theta^D \theta^B \theta}}^C
= j \theta^A \theta^B \theta^C \theta^D ;
\end{equation}
\indent
therefore, as $1-j \neq 0$, one has $\theta^A \theta^B \theta^C \theta^D = 0$.
\newline
\indent
The dimension of this $Z_3$-graded generalization of  Grassmann algebra
is equal to $ N + N^2 + (N^3 - N)/3$; we may also add a "neutral" element
denoted by {\tbf 1} and commuting with all other generators.
\newline
\indent
One can note a dissymetry between the components of this algebra with the
grades 1 et 2 : as a matter of fact, there are $N$ elements of grade 1, 
(the $\theta$'s ) and $N^2$ elements of grade 2 ($\theta \theta$).
\newline
\indent
A natural way to re-establish the symmetry is to introduce the set of $N$
"{\it conjugate}" generators , ${\bar \theta}^A$, of grade 2, that would
satisfy conjugate ternary relations (in which $j$ is replaced by $j^2$):
\begin{equation}
{\bar \theta}^A {\bar \theta}^B {\bar \theta}^C = j^2 {\bar \theta}^B
{\bar \theta}^C {\bar \theta}^A
\end{equation}
\indent
The ternary relation between the $\theta^A$'s can be interpreted as follows:
\newline
$\theta^A {\underbrace{\theta^B \theta}}^C =
j {\underbrace{\theta^B \theta}}^C \theta^A$, which suggests
the following relations between the generators $\theta^A$ and ${\bar \theta}^B$ :
\begin{equation}
\theta^A {\bar \theta}^B  = j {\bar \theta}^B \theta^A , \   \
{\bar \theta}^B \theta^A = j^2 \theta^A {\bar \theta}^B .
\end{equation}
\indent
The $Z_3$-graded algebra so defined can be naturally divided in three parts, 
of grade 0, 1 et 2 respectively, with the dimensions of the sub-spaces of 
grades 1 and 2 being equal: one can write symbolically $A = A_0 + A_1 + A_2$, 
where
\vskip 0.2cm
\hskip 0.4cm
$A_0$ contains: 
{ {\tbf 1}, $\theta^A {\bar \theta}^B , \theta^A \theta^B
\theta^C , {\bar \theta}^A {\bar \theta}^B {\bar \theta}^C , \theta^A \theta^B
{\bar \theta}^C {\bar \theta}^D \ \ {\rm and} \ \ \theta^A \theta^B \theta^C
{\bar \theta}^D {\bar \theta}^E {\bar \theta}^F$ ;
\vskip 0.2cm
\hskip 1cm $A_1$ contains: 
$\theta^A, {\bar \theta}^B {\bar \theta}^C , \theta^A \theta^B
{\bar \theta}^C, \theta^A {\bar \theta}^A {\bar \theta}^B {\bar \theta}^C $,
\vskip 0.2cm
\hskip 0.5cm and $A_2$ contains: ${\bar \theta}^A, \theta^A \theta^B,
\theta^A {\bar \theta}^B {\bar \theta}^C, \theta^A \theta^B \theta^C
{\bar \theta}^D$ .
\vskip 0.2cm
\indent
In the case of usual $Z_2$-graded Grassmann algebras the anti-commutation
between the generators of the algebra and the assumed associativity imply
automatically the fact that {\it all} grade $0$ elements {\it commute} with
the rest of the algebra, while {\it any two} elements of grade $1$ anti-commute.
\newline
\indent
In the case of the $Z_3$-graded generalization such an extension of ternary
and binary relations {\it does not follow automatically}, and must be imposed
explicitly. If we decide to extend these relations to {\it all} elements of
the algebra having a well-defined grade (i.e. the monomials in $\theta$'s
and $\bar{\theta}$'s , then many additional expressions must vanish, e.g.:
\vskip 0.15cm
\hskip 2cm
$\theta^A {\underbrace{\theta^B {\bar \theta}}}^C = 
{\underbrace{\theta^B {\bar \theta}}}^C \theta^A =
\theta^B {\underbrace{{\bar \theta}^C \theta}}^A = 
{\bar \theta}^C \theta^A \theta^B = 0$ ;
\vskip 0.2cm
because on the one side, $\theta^A {\bar \theta}^C$ is of grade 0 and commutes 
with all other elements; at the same time, commuting ${\bar \theta}^C$ with
$\theta^A \theta^B$ one gets twice the factor $j^2$, which leads to the
overall factor $ j {\bar \theta}^C \theta^A \theta^B $; this produces a
contradiction which can be solved only by supposing that 
$\theta^A \theta^B {\bar \theta}^C = 0$. 
\newline
\indent
The resulting $Z_3$-graded algebra contains only the following products of
generators:
\begin{equation}
A_1 =  \theta , \lbrace {\bar \theta} {\bar \theta} \rbrace ; \ \ \ \  
A_2 =  {\bar \theta}, \  \ \lbrace \theta \theta \rbrace ; \  \ \  \ 
A_0 = \lbrace \theta {\bar \theta} \rbrace , 
\ \ \lbrace \theta \theta \theta \rbrace , \ \ 
\lbrace {\bar \theta} {\bar \theta} {\bar \theta} \rbrace
\end{equation}
\indent
Let us note that the set of grade $0$ (which obviously forms a sub-algebra
of the $Z_3$-graded Grassmann algebra) contains the products which could
symbolize the only observable combinations of {\it quark fields} in quantum 
chromodynamics based on the $SU(3)$-symmetry.
\newline
\indent
If we align the basis of our algebra, with all the elements of grade $0$ first,
next all the elements of grade $1$ and finally the elements of grade $2$ in
a one-column vector, a general linear transformation that would leave these 
entries in the same order can be symbolized by a matrix whose entries have 
a definite $Z_3$-grade placed as follows:
\begin{equation}
\pmatrix{0 & 2 & 1 \cr 1& 0 & 2 \cr 2 & 1 & 0}  \pmatrix{  0 \cr  1 \cr  2 }
= \pmatrix{0 \cr 1 \cr 2}
\end{equation}
\indent
Note that the position of the three grades did not change in the resulting 
column; we shall call such an operator a {\it grade 0} matrix. 
We can introduce two other kinds of matrices that raise all the grades 
by $1$ (resp. by $2$), and call them respectively {\it grade 1} and 
{\it grade 2} matrices, as follows:
\begin{equation}
\pmatrix{1 & 0 & 2 \cr 2 & 1 & 0 \cr 0 & 2 & 1} \pmatrix{0 \cr 1 \cr 2} = 
\pmatrix{1 \cr 2 \cr 0}, \ \ {\rm and} \ \ \pmatrix{2 & 1 & 0 \cr 0 & 2 & 1 
\cr 1 & 0 & 2} \pmatrix{0 \cr 1 \cr 2} = \pmatrix{2 \cr 0 \cr 1}
\end{equation}
\indent (the numbers $0, 1, 2$ symbolize the grades of the respective
entries in the matrices).
\newline
\indent
If we restrict the character of the matrices, admitting only complex-valued
matrix elements, then the grades $0, 1$ and $2$ will reduce themselves to
the following three types of $3 \times 3$-block matrices:
\begin{equation}
\pmatrix{a & 0 &0 \cr 0 & b& 0 \cr 0 & 0 &c} , \  \ \  \ 
\pmatrix{0 & \alpha & 0 \cr 0 & 0 & \beta \cr \gamma & 0 &0 } , 
\  \  \  \  \pmatrix {0 & 0 & \gamma \cr \alpha & 0& 0 \cr 0 & \beta & 0 }
\end{equation}
\indent
representing arbitrary matrices with respective grade $0, 1$ and $2$.
\newline
\indent
It is easy to check that these grades add up modulo $3$ under the associative
matrix multiplication law.
\newline
\indent
Let $B, C$ denote two matrices whose grades are $grad(A)=a$ and $grad(B)=b$ ,
respectively. We can define the {\it $Z_3$-graded commutator} $[B,C]$ as
follows:
\begin{equation}
[ B , C ]_{Z_3} := B C - j^{bc} C B ,
\end{equation}
\indent
(Note that this $Z_3$-graded commutator does not satisfy the Jacobi identity).
Let $\eta$ be a matrix of grade $0$; we can choose for the sake of simplicity
\begin{equation}
\eta = \pmatrix{0 & 1 & 0 \cr 0 & 0 & 1 \cr 1 & 0 & 0}
\end{equation}
\newline
\indent
With the help of the matrix $\eta$ we can define a formal "differential" on
the $Z_3$-graded algebra of $3 \time 3$ matrices as follows:
\begin{equation}
d B := [ \eta , B ]_{Z_3} = \eta B - j^b B \eta
\end{equation}
\indent
It is easy to show that $d (B C ) = (d B) C + j^b B (d C) $ and that $d^3 = 0$.
The first identity is trivial, whereas the last one follows from the fact that
$\eta^3 = Id$ does commute with all the elements of the algebra.
\newline
\indent
A formal algebraic analogue of connection and curvature forms have been
discussed elsewhere (cf. ref. [9] and [10]). Here we shall show how such a 
$Z_3$-graded exterior differential may be realized on a differential manifold.
Some interesting ideas concerning similar techniques may be found in ref.[11].
\vskip 0.6cm
{\tbf 3. $Z_3$-GRADED  EXTERIOR  DIFFERENTIAL.}
\vskip 0.4cm
\indent
Let $M_n$ be a differentiable manifold of dimension $N$, with local coordinates 
${x^k}$. The variables $x^k$ commute and with respect to the $Z_3$-grading 
are of grade $0$. Our aim now is to define such a linear operation acting
on functions of the coordinates $x^k$, and by extension, on a larger algebra
of forms that still remains to be defined, that would reproduce the properties
of the "algebraic differential" introduced above. To do this, we have to
define a linear operation $d$ which acts on the functions of $x^k$; for the
definition, it is enough to define its action on the coordinates $x^k$ and
on their products.
\newline
\indent
We postulate that the linear operator $d$ applied to $x^k$ produces a 1-form 
whose $Z_3$-grade is $1$ by definition; when applied two times by iteration, 
it will produce a new entity, which we shall call a {\it 1-form of grade $2$,} 
denoted by $d^2 x^k$. Finally, we require that $d^3 = 0$. 
\newline
\indent
Let {\it F}({\it M}) denote the algebra of functions $C^{\infty} (M)$, 
over which the $Z_3$-graded algebra generated by the forms $dx^i$ and $d^2 x^k$ 
behaves as a {\it left} module. In other words, we shall be able to multiply
the forms $dx^i$ , $d^2 x^k$, $dx^i dx^k$ , etc. by the functions {\it on the
left} only; right multiplication will just not be considered here.
That is why we will write by definition, e.g.                 
\begin{equation}
d (x^i x^k) := x^i dx^k + x^k dx^i
\end{equation}
\indent
We shall also assume the following Leibniz rule for the operator $d$ with
respect to the multiplication of the $Z_3$-graded forms: when $d$ crosses
a form of grade $p$, and of arbitrary {\it rank}, the factor $j^p$ appears 
as follows:  
\begin{equation}
d( \omega \ \ \phi ) = ( d \omega) \phi + j^p \omega ( d \phi)
\end{equation}
\indent
Let us note that in contrast with the $Z_2$-graded case, the forms are treated
as one whole, even when multiplied from the left by an arbitrary function;
that means that we can not identify  e.g. 
\newline
\centerline{$(\omega_i dx^i)(\phi_k dx^k)$ with $(\omega_i \phi_k) dx^i dx^k$} 
\newline
\indent
This is equivalent with saying that the products of functions by the forms 
are to be understood in the sense of tensor products, which is associative, 
but non-commutative.
\newline
\indent
Nevertheless, we shall show later that such an identification can be done
for the forms of maximal degree (i.e. $3$), which contain the products of
the type $ d x^i d x^k d x^m$ or $d x^i d^2 x^m$, whose exterior differentials
vanish irrespective of the order of the multiplication; as a matter of fact,
it will be easy to show that with the rules of differentiation that we expose
below,
\begin{equation}
d ((\alpha_i d x^i)(\beta_k d x^k)(\gamma_m d x^m)) = d ((\alpha_i \beta_k 
\gamma_m) d x^i d x^k d x^m)) = 0 .
\end{equation}
\indent
With the so established $Z_3$-graded Leibniz rule, the postulate $d^3 = 0$ 
suggests in an almost unique way the ternary and binary commutation rules
for the differentials $dx^i$ and $d^2 x^k$. To begin with, consider the
differentials of a function of the coordinates $x^k$, with the "first
differential"  $df$ coinciding with the usual one:
\begin{equation}
df :=  (\partial_i f) dx^i \ \ ; \ \ \  \ d^2 f := (\partial_k \partial_i f) dx^k dx^i
+ (\partial_i f) d^2 x^i \ \ ;
\end{equation}
\begin{equation}
d^3 f = (\partial_m \partial_k \partial_i f) dx^m dx^k dx^i + (\partial_k
\partial_i f) d^2 x^k dx^i + j(\partial_k \partial_i f) dx^i d^2 x^k +
(\partial_k \partial_i f) dx^k d^2 x^i ;
\end{equation}
(we remind that the last part of the differential, $(\partial_i f)d^3 x^i$, 
vanishes by virtue of the postulate $d^3 x^i = 0$).
\newline
\indent
Supposing that the partial derivatives commute, exchanging the summation
indices $i$ et $k$ in the last expression and replacing $1 + j$ by $- j^2$, 
we arrive at the following two conditions that lead to the vanishing of $d^3 f$ :
\begin{equation}
dx^m dx^k dx^i + dx^k dx^i dx^m + dx^i dx^m dx^k = 0 \ \ ; \  \ \  \
d^2 x^k d x^i - j^2  d x^i d^2 x^k = 0 \ \ .
\end{equation}
\indent
which lead in turn to the following choice of relations:
\begin{equation}
d x^i d x^k d x^m = j d x^k d x^m d x^i , \  \ {\rm and} \  \  d x^i d^2 x^k = 
j d^2 x^k d x^i .
\end{equation}
\indent
By extending these rules to {\it all} the expressions with a well-defined
grade, and applying the associativity of the $Z_3$-exterior product, we
see that all the expressions of the type $d x^i d x^k d x^m d x^n$ and 
$d x^i d x^k d^2 x^m$ must vanish, and along with them, also the monomials 
of higher order that would contain them as factors.
\newline
\indent
Still, this is not sufficient in order to satisfy the rule $d^3 = 0$ on all 
the forms spanned by the generators $d x^1$ and $d^2 x^k$. It can be proved
without much pain that the expressions containing $d^2 x^i d^2 x^k$ must
vanish, too. For example, if we take the particular 1-form $x^i d x^k$ and
and apply to it the operator $d$, we get
\begin{equation}
d (x^i d x^k) = d x^i d x^k + x^i d^2 x^k;  
\end{equation}
\begin{equation}
d^2 (x^i d x^k) = d^2 x^i d x^k + (1 + j) d x^i d^2 x^k = d^2 x^i d x^k - d^2 x^k d x^i ;
\end{equation}
\indent
which leads to $d^3 (x^i d x^k) = d^2 x^i d^2 x^k - d^2 x^k d^2 x^i $; then,
if we want to keep both the associativity of the "exterior product" and the
ternary rule for the entities of grade $2$, i.e. 
$d^2 x^i d^2 x^k d^2 x^m = j^2 d^2 x^k d^2 x^m d^2 x^i$, then the only
solution is to impose $d^2 x^i d^2 x^k = 0$ and to set forward the additional
rule declaring that any expression containing {\it four or more} operators
$d$ must vanish identically.
\newline
\indent
With this set of rules we can check that $d^3 = 0$ on all the forms, whatever
their grade or degree. Let us show how such calculus works on the example of
a 1-form $\omega = \omega_k d x^k$:
\begin{equation}
d(\omega_k d x^k) = (\partial_i \omega_k) d x^i d x^k + \omega_k d^2 x^k ;
\end{equation}
\begin{equation}
d^2 (\omega_k d x^k) = (\partial_m \partial_i \omega_k) d x^m d x^i d x^k +
(\partial_i \omega_k) (d^2 x^i d x^k + j d x^i d^2 x^k) + \partial_i \omega_k
d x^i d^2 x^k ;
\end{equation}
after exchanging the summation indices $i$ and $k$ in two last terms and
using the fact that $j + 1 = - j^2$ and the commutation relations between
$d x^k$ and $d^2 x^i$, we can write
\begin{equation}
d^2 (\omega_k d x^k) = (\partial_m \partial_i \omega_k) d x^m d x^i d x^k 
+ (\partial_i \omega_k - \partial_k \omega_i) d^2 x^i d x^k .
\end{equation}
where it is interesting to note how the usual anti-symmetric exterior
differential appears as a part of the whole expression.
\newline
\indent
It is also easy to check that
\begin{equation}
Im (d) \subseteq Ker(d^2) , \  \  {\rm and} \  \  Im (d^2) \subseteq Ker (d)
\end{equation}

\vskip 0.6cm
{\tbf 4.  COVARIANT  DIFFERENTIAL, CONNECTION  AND  CURVATURE.}
\vskip 0.4cm
\indent
Let {\cal A} be an associative algebra with unit element, and let {\cal H} be
a free left module over this algebra. Let $A$ be a {\cal A}-valued 1-form 
defined on a differential manifold $M$, and let $\Phi$ be a function on the 
manifold $M$ with values in the module {\cal H}. 
\newline
\indent
We shall introduce the {\it covariant differential} as usual:
\begin{equation}
D \Phi := d \Phi + A  \Phi  ;
\end{equation}
\indent
If the module is a free one, any of its elements $\Phi$ can be represented
by an appropriate element of the algebra acting on a fixed element of {\cal H}, 
so that one can always write $ \Phi = B \Phi_o $; then the action of the
group of automorphisms of {\cal H} can be translated as the action of the 
same group on the algebra {\cal A}.
\newline
\indent
Let $U$ be a function defined on $M$ with its values in the group of the 
automorphisms of {\cal H}. 
The definition of a covariant differential is equivalent with the requirement
$D U^{-1} B = U^{-1} D B $; as in the usual case, this leads to the following
well-known transformation for the connection 1-form $A$ :
\begin{equation}
A \Rightarrow U^{-1} A U + U^{-1} d U  ;
\end{equation}
\indent
But here, unlike in the usual theory, the second covariant differential
$D^2 \Phi$ is not an automorphism: as a matter of fact, we have:
\begin{equation}
D^2 \Phi = d ( d \Phi + A \Phi) + A (d \Phi + A \Phi) = 
d^2 \Phi + d A \Phi + j A d \Phi + A d \Phi + A^2 \Phi  ;
\end{equation}
the expression containing $d^2 \Phi$ and $d \Phi$ ; whereas $D^3 \Phi$ is an
automorphism indeed, because it contains only $\Phi$ multiplied on the left
by an algebra-valued 3-form:
\begin{equation}
D^3 \Phi = d (D^2 \Phi) + A (D^2 \Phi) , 
\end{equation}
which gives explicitly:
\begin{equation}
d( d^2 \Phi + dA \Phi + j A d \Phi + A^2 \Phi ) + A ( d^2 \Phi + 
d A \Phi + j A d \Phi + A d \Phi + A^2 \Phi)
\end{equation}
With a direct calculus one observes that all the terms containing
$d \Phi$ or $d^2 \Phi$ simplify because of the identity $1+j+j^2=0$, leaving
only
\begin{equation}
D^3 \Phi = ( d^2 A + d ( A^2 ) + A d A + A^3 ) \Phi = (D^2 A) \Phi := \Omega \Phi ;
\end{equation}
\indent
Obviously, because $D (U^{-1} \Phi) = U^{-1} (D \Phi) $, one also has: 
\vskip 0.2cm
\centerline{$D^3 (U^{-1} \Phi) = U^{-1} ( D^3 \Phi ) = U^{-1} \Omega \Phi = 
U^{-1} \Omega U U^{-1} \Phi$,} 
\vskip 0.2cm
\indent
which proves that the 3-form $\Omega$ transforms as usual, 
$\Omega \Rightarrow U^{-1} \Omega U$ when the connection 1-form
transforms according to the law:
$A \Rightarrow U^{-1} A U + U^{-1} dU $.
\newline
\indent
It can be also proved by a direct calculus that the curvature 3-form $\Omega$
does vanish identically for $A = U^{-1} dU $. This computation illustrates
very well the technique of the $Z_3$-graded exterior differential calculus
introduced above: as a matter of fact, one has
\begin{equation}
d( U^{-1} dU ) = d U^{-1} dU + U^{-1} d^2 U ,
\end{equation}
\indent
so that the term corresponding to $d^2 A$ gives:
\begin{equation}
d^2 ( U^{-1} dU ) = d^2 U^{-1} d U + j d U^{-1} d^2 U + d U^{-1} d^2 U; 
\end{equation}
\indent
next, the term corresponding to $d(A^2) = d(U^{-1} dU U^{-1} dU ) $ gives 
\begin{equation}
d U^{-1} dU U^{-1} dU + U^{-1} d^2 U U^{-1} dU +
j U^{-1} dU d U^{-1} dU + j U^{-1} dU U^{-1} d^2 U ;
\end{equation}
\indent
whereas
\begin{equation}
A dA = U^{-1} dU d U^{-1} dU + U^{-1} dU U^{-1} d^2 U ;
\end{equation}
and finally, the term corresponding to $A^3 = U^{-1} dU U^{-1} dU U^{-1} dU$
can be written as $ - d U^{-1} d U U^{-1} dU $ by virtue of the identity
$dU U^{-1} = - U dU^{-1}$ which follows from the Leibniz rule applied to 
$U U^{-1} = {\tbf 1}$, i.e. $d(U U^{-1}) = dU U^{-1} + U d U^{-1} = 0.$ 
\vskip 0.3cm
Using this identity whenever possible, and replacing $1 + j$ by $- j^2$,
we can reduce the whole expression to the following sum of three terms
\begin{equation}
d^2 U^{-1} d U + U^{-1} d^2 U U^{-1} dU - j^2 U^{-1} dU dU^{-1} dU
\end{equation}
whose vanishing does not at all seem obvious.
\newline
\indent
However, it is not very difficult to prove that this expression is identically
null. First of all, it is enough to prove the vanishing of the expression
\newline
\indent
\hskip 1.5cm
$d^2 U^{-1} + U^{-1} d^2 U U^{-1} - j^2 U^{-1} dU d U^{-1} $ ,
\newline
because all the three terms contain the same factor $d U$ on the right; then,
by multiplying on the left by $U$, we get
\begin{equation}
U d^2 U^{-1} + d^2 U U^{-1} - j^2 dU dU^{-1}
\end{equation}
\indent
At this point let us note that $d^2 (U U^{-1}) = d^2 (Id) = 0$, but then,
according to our $Z_3$-graded Leibniz rule,
\begin{equation}
d^2(U U^{-1}) = d (dU U^{-1} + U dU^{-1}) = d^2 U U^{-1} + j dU dU^{-1} +
dU dU^{-1} + U d^2 U^{-1} ;
\end{equation}
so that  $ U d^2 U^{-1} + d^2 U U^{-1} = - dU dU^{-1} - j dU dU^{-1} $, and
the result can be written as
\begin{equation}
-dU dU^{-1} - j dU dU^{-1} - j^2 dU dU^{-1} = 0
\end{equation}
\indent
because here again the common factor $ - dU dU^{-1}$ is multiplied by 
$(1 +j +j^2)= 0$ , which completes the proof.
\vskip 0.6cm
{\tbf 5. EXPRESSIONS  IN  LOCAL  COORDINATES.}
\vskip 0.4cm
\indent
The curvature 3-form $\Omega = d^2 A + d (A^2) + A dA + A^3$ is of grade $0$; 
therefore it must be decomposed along the elements $dx^i dx^k dx^m$ and
$d^2 x^i dx^k$. 
Here is how we can compute its components in a local coordinate system. 
By definition,  $A = A_i dx^i$, so we have:
\begin{equation}
dA = \partial_i A_k dx^i dx^k + A_k d^2 x^k ;  
\end{equation}
\begin{equation}
d^2 A = \partial_m
\partial_i A_k dx^m dx^i dx^k + \partial_i A_k d^2 x^i dx^k + j \partial_i
A_k dx^i d^2 x^k + \partial_i A_k dx^i d^2 x^k ;
\end{equation}
\indent
After replacing $1 + j$ by $ - j^2$, and taking into account the relation 
$ dx^k d^2 x^i = j d^2 x^i dx^k $, we get:
\begin{equation}
d^2 A = ( \partial_m \partial_i A_k ) dx^m dx^i dx^k + (\partial_i A_k -
\partial_k A_i ) d^2 x^i d x^k ;
\end{equation}
\begin{equation}
{\rm Then,} \  \ d (A^2) + A d A = d A A + j A dA + A dA = dA A - j^2 A dA , 
\end{equation}
\indent
which leads easily to
\begin{equation}
(\partial_i A_k A_m) dx^i dx^k dx^m - j^2 (A_m \partial_i A_k) dx^m dx^i dx^k
+ A_k A_m d^2 x^k dx^m - j^2 A_m A_k dx^m d^2 x^k ;
\end{equation}
\indent
and due to the relations $dx^m d^2 x^k = j d^2 x^k dx^m$ et $dx^m dx^i dx^k = 
j dx^i dx^k dx^m$,
\begin{equation}
d (A^2) + A dA = (A_m \partial_i A_k - \partial_i A_k A_m) dx^m dx^i dx^k +
(A_k A_m - A_m A_k) d^2 x^k dx^m .
\end{equation}
\indent
Finally, as $A^3 = A_i A_k A_m dx^i dx^k dx^m $, the curvature 3-form can
be written in local coordinates as follows:
\begin{equation}
\Omega = d^2 A + d (A^2) + A dA + A^3 = \Omega_{ikm} dx^i dx^k dx^m + 
F_{ik} d^2x^i dx^k
\end{equation}
\begin{equation}
{\rm where} \ \ \Omega_{ikm} := \partial_i \partial_k A_m + A_i \partial_k A_m - 
\partial_k A_m A_i + A_i A_k A_m ,
\end{equation}
\begin{equation}
{\rm and} \  \ F_{ik} := \partial_i A_k - \partial_k A_i + A_i A_k - A_k A_i ;
\end{equation}
In $F_{ik}$ one can easily recognize the 2-form of curvature of the usual gauge
theories.
\newline
\indent
We know that the expression $F_{ik}$ is covariant with respect to the gauge
transformations; on the other hand, the 3-form $\Omega$ is also covariant;
therefore, the local expression $\Omega_{ijk}$ must be covariant, too.
As a matter of fact, it can be expressed as a combination of covariant
derivatives of the 2-form $F_{ik}$. 
\newline
\indent
In order to find the covariant expression of $\Omega_{ikm}$, it suffices to
recall that due to the particular symmetry of the ternary exterior product
$dx^i dx^k dx^m $,
we can replace $\Omega_{ikm}$ by $\frac{1}{3} (\Omega_{ikm} + j^2  \Omega_{kmi} 
+ j \Omega_{mik})$  and analyze the {\it abelian} case, when this expression
reduces itself to $\Omega_{ikm} = \partial_i \partial_k A_m$.
\vskip 0.2cm
\indent
Substituting for $\partial_i \partial_k A_m$ the equivalent expression
$\frac{1}{3}(\partial_i \partial_k A_m + j^2 \partial_k \partial_m A_i +
j \partial_m \partial_i A_k )$
\vskip 0.2cm
and then 
$\frac{1}{3} ( j (\partial_k \partial_m A_i - \partial_i \partial_k A_m) +
j ( \partial_m \partial_i A_k - \partial_i \partial_k A_m ) ) $ , because
$ 1 = -j - j^2$,
\vskip 0.2cm
we can easily recognize
$\frac{1}{3} ( j \partial_i [ \partial_m A_k - \partial_k A_m] + 
j^2 \partial_k [ \partial_m A_i - \partial_i A_k] )$ ;
\newline
\indent
which in a general non-abelian case must lead to the following expression:
\begin{equation}
\Omega_{ikm} =  \frac{1}{3} [ j D_i F_{mk} + j^2 D_k F_{mi} ] ,
\end{equation}
\indent
or, equivalently,
\begin{equation}
\Omega_{ikm} = - \frac{1}{6} [ D_i F_{mk} + D_k F_{mi} ] + 
\frac{ i \sqrt{3}}{6} [D_i F_{mk} - D_k F_{mi} ]
\end{equation}

\vskip 0.6cm
{\tbf 6.  DUALITY, INTEGRATION,  VARIATIONAL  PRINCIPLE.} 
\vskip 0.4cm
\indent
The natural symmetry between $j$ et $j^2$ , which leads to the possibility
of choosing one of these two complex numbers as the generator of the group
$Z_3$ , and simultaneous interchanging the r\^oles between the grades $1$
and $2$, suggests that we could extend the notion of complex conjugation 
$j \Rightarrow (j)^* :=j^2 $, with $((j)^*)^* =j$, to the algebra of $Z_3$-
graded exterior forms and the operator $d$ itself.
\newline
\indent
It does not seem reasonable to use the "second differentials" $d^2 x^i$ as the 
objects conjugate to the "first differentials"  $d x^i$, because the rules
of $Z_3$-graded exterior differentiation we have imposed break the symmetry
between these two kinds of differentials: remember that the products 
$d x^i d x^k$, and $d x^i d x^k d x^m$ are admitted, while we require that 
$d^2 x^i d^2 x^k$ and $d^2 x^i d^2 x^k d^2 x^m $ must vanish.
\newline
\indent
This suggests the introduction of a "{\it conjugate}" differential $\delta$
of grade $2$, the image of the differential $d$ under the conjugation $*$, 
satisfying the following conjugate relations:
\begin{equation}
\delta x^i \delta x^k \delta x^m = j^2 \delta x^k \delta x^m \delta x^i , \ \ 
\delta x^i \delta^2 x^k = j^2 \delta^2 x^k \delta x^i \ \ .
\end{equation}
\indent
One notes that $\delta^2 x^k$ is of grade 1 ($2+2 = 4 = 1 \ \ (mod \ \ 3)$).
\newline
\indent
All the relations existing between the operator $d$ and the exterior forms 
generated by $d x^i$ and $d^2 x^k$ are faithfully reproduced under the
conjugation $*$ if we consider the $Z_3$-graded algebra generated by the
entities $\delta x^i$ and $\delta^2 x^k$ as a {\it right module} over the
algebra of functions {\it F}({\it M}) , with the operator $\delta$ acting  
{\it on the right} on this module.
\newline
\indent
The rules $d^3 = 0$ and $\delta^3 = 0$ suggest their natural extension:
\begin{equation}
d  \delta = \delta  d  = 0
\end{equation}
\indent
We would like to be able to form quadratic expressions that could define a
scalar product; to do this, we should postulate that the algebra generated by
the elements $d x^i, d^2 x^k$ and its conjugate algebra generated by the 
elements $\delta x^i , \delta x^k$ commute with each other.
\newline
\indent
Then, we can define scalar products for the forms of maximal degree $3$:
$ < \omega \mid \phi > := * \omega \phi $ , and integrating this result with
respect to the usual volume element defined on the manifold $M$, which gives
explicitly:
\begin{equation}
\int {\bar \omega_{ikm}} \phi_{prs} < \delta x^i \delta x^k \delta x^m \mid 
d x^p d x^r d x^s >  ,  \int {\bar \psi_{ik}} \chi_{mn} 
< \delta x^i \delta^2 x^k \mid d^2 x^m d x^n > 
\end{equation}
\indent
What remains now is to determine the scalar products of the basis of forms;
in order to assure the hermiticity of the product, one can always choose an
"orthonormal" basis in which we should have:
\begin{equation}
< \delta x^i \delta x^k \delta x^m \mid d x^p d x^r d x^s > = \delta^i_s
\delta^k_r \delta^m_p \ \ {\rm et} \ \ < \delta x^i \delta^2 x^k \mid
d^2 x^m d x^n > = \mu \delta^i_n \delta^k_m .
\end{equation}
\indent
Here the first scalar product is normed to $1$, and $\mu$ is the ratio between  
the two types of "elementary volume".
\newline
\indent
We shall consider the two types of forms of degree $3$, $d x^i d x^k d x^m$ 
and $d^2 x^p d x^r$, as being mutually orthogonal.
\newline
\indent
Because in both cases the differentials of the volume forms are identically
null, we can formally apply Stokes' formula with a vanishing contribution
of the "surface term", which enables us to derive the classical Euler-Lagrange
equations. With the Lagrangean density defined on the manifold as 
$<\Omega \mid \Omega>$ we get:
\begin{equation}
L = [\frac{4}{3} D_i F_{mk} D^i F^{mk} - \frac{2}{3} D_k F_{mi} D^i F^{mk} ]
+ \mu [ 4 F_{ik} F^{ik} ]
\end{equation}
\indent
(We suppose here that the manifold $M$ is flat, and we raise and lower the
indeces by means of the metric tensor $\delta_{ik}$ and its inverse $
\delta^{kl}$).
\newline
\indent
The corresponding Euler-Lagrange are then easily deduced:
\begin{equation}
D^m D^i D_i F_{mk} -  D^i D^k D_m F_{ik} + \frac{3 \mu}{4}  ( D^i F_{ik}) = 0
\end{equation}
In the abelian case, and with the usual choice of gauge ($ \partial^i A_i = 0$),
these equations reduce themselves to:
\begin{equation}
\bigtriangleup \bigtriangleup A_k + \frac{3 \mu}{4} \bigtriangleup A_k =  0
\end{equation}
\newline
\indent
It is worthwhile to note that this scheme can be readily generalized to a
$Z_N$-graded case, with an exterior differential operator satisfying $d^N = 0$,
leading to the invariants built with the higher-order covariant derivatives 
of the usual curvature tensor, leading to the equations containing higher
powers of the Laplace-Beltrami operator.
\newline
\indent
The equations of this type have been studied before ([L.Vainerman, (1974)])
In quantum field theory, they appear in the perturbative developments of
non-linear effective Lagrangeans (the so-called Schwinger terms). In such 
expansions, the consecutive powers of the Laplacian (or the d'Alembertian) 
operator appear, with the coeficients depending on the particular choice of 
Lagrangean. It would be interesting to compare the coefficients appearing in 
these theories with the coefficients depending on the choice of normalization 
of our various "volume elements" of higher orders, which appear in the 
above model.
\newline
\indent
Nevertheless, from the mathematical point of view it seems that a theory
based on the $Z_3$-grading is particularly interesting because of its
exceptional character related to the fact that the group $S_3$ is the 
{\it last} of the permutation groups that has a {\it faithful} representation 
in complex plane, which could lead to some special applications in physics, 
e.g. a better description of the quark fields ( cf.[R.Kerner, 1992], 
[B. Le Roy, 1995], and [ V.Abramov, R.Kerner, B.Le Roy (1995)].).
\newline
\indent
Another possible application of this formalism may be a new description
of differential operators on Riemannian manifolds. Consider a linear operation
$d$ that produces a {\it symmetric} $2$-form from a given $1$-form,
\vskip 0.2cm
\centerline{$(d \xi)_{ij} = \partial_i \xi_j + \partial_j \xi_i$}
\vskip 0.2cm
\indent
The complex thus includes $1$-forms and symmetric $2$-forms (metrics). Now,
let $d$ of a metric be defined as its Christoffel connection, and $d$ of
the latter as Riemannian curvature. Then we have $d^2 \neq 0$, but $d^3 =0$.
\newline
\indent
The applications to gauge theories with well defined non-abelian gauge
groups will be the object of our forthcoming articles.

\newpage
\indent
\vskip 0.3cm
\indent
{\tbf R\'ef\'erences}
\vskip 0.2cm
\indent
1. M. Dubois-Violette, R. Kerner, J. Madore, Journ.Math.Phys.{\tbf 31} p.316,  
\newline
\indent
\hskip 0.5cm
{\it ibid}, p. 323 (1990)
\newline
\indent
2. M. Dubois-Violette, J. Madore, R. Kerner, Class. and Quantum Gravity,
\newline
\indent
\hskip 0.5cm
{\tbf 8}, p. 1077 - 1089  (1991)
\newline 
\indent
3. A. Connes, J. Lott, Nucl. Phys. B (Proc.Suppl.) {\tbf 18} , p.29 (1990)
\newline
\indent
4. R. Coquereaux, G. Esposito-Farese, P. Vaillant, Nucl. Phys. B, {\tbf 353} 
\newline
\indent
\hskip 0.5cm
p. 689 (1991)
\newline
\indent
5. R. Kerner, Journal of Geometry and Physics, {\tbf 11} (1 - 4) p.325 (1993)
\newline
\indent
6. A.H. Chamseddine, G. Felder, J. Frohlich, Phys. Lett. B  {\tbf 296}, 
\newline
\indent
\hskip 0.5cm
p. 109 (1992)
\newline 
\indent
7. V. Abramov, R. Kerner, B. Le Roy, to appear (1995)
\newline
\indent
8. R. Kerner, C.R.Acad.Sci., Paris, {\tbf 312} (Ser. II) , p.191-196, (1991) 
\newline
\indent
9. R. Kerner, Journal of Math. Physics, {\tbf 33} (1), p. 411 - 416, (1992)
\newline
\indent
10. R. Kerner, in "Generalized Symmetries in Physics", H.-D. Doebner, 
\newline
\indent 
\hskip 0.5cm
V.K. Dobrev and A.G. Ushveridze ed., World Scientific, p.375 - 394, (1994)
\newline
\indent
11.V.Takhtajan, Comm.Math.Phys., {\tbf 160}, p. 295 (1994)
\newline
\indent
12. B. Le Roy, C.R.Acad.Sci., to appear (1995)
\newline
\indent
13. L. Vainerman, Soviet.Math.Dokl,  (1974)

\vskip 0.2cm
{\it Acknowledgements} .  
Numerous enlightening discussions  with V. Abramov, B. Le Roy, L. Vainerman, 
M. Dubois-Violette and J. Madore are gratefully acknowledged.

\vskip 0.2cm
\centerline{\it Laboratoire GCR,} 
\centerline{\it Universit\'e Pierre-et-Marie-Curie - CNRS URA 769,}
\centerline{\it Tour 22, Bo{\'i}te 142, 4, Place Jussieu, 75005 Paris.}
\centerline{{\it e-mail address} : rk@ccr.jussieu.fr} 
\centerline{{\it FAX} : (33 1) 44 27 72 87 .}

\end{document}